\documentclass[useAMS,usenatbib]{mn2e}
\newcommand{\teff}{\mbox{$T_{\rm eff}$}}
\newcommand{\logg}{\mbox{$\log g$}}

\newcommand{\vsini}{\mbox{$v \sin i$}}
\newcommand{\mactrb}{\mbox{$v_{\rm mac}$}}
\newcommand{\mictrb}{\mbox{$v_{\rm mic}$}}
\newcommand{\nusini}{\mbox{$\delta \nu_{\rm s} \sin i$}}
\newcommand{\loggf}{\mbox{$\log gf$}}
\newcommand{\kms}{\mbox{km\,s$^{-1}$}}
\usepackage[dvips]{graphicx}
\usepackage{amsmath}
\usepackage{latexsym}
\usepackage{amsfonts}
\usepackage{amssymb}
\usepackage{amsmath}
\usepackage{lscape}
\title[Determining stellar macroturbulence]
 {Determining stellar macroturbulence using asteroseismic rotational velocities from {\it Kepler}}

\author[A.P.~Doyle et al.] {Amanda P.~Doyle$^{1}$\thanks{E-mail: a.doyle@keele.ac.uk}, Guy R.~Davies$^{2,3}$, Barry~Smalley$^{1}$, William J.~Chaplin$^{2,3}$,\newauthor  Yvonne~Elsworth$^{2,3}$\\
$^{1}$ Astrophysics Group, Keele University, Staffordshire ST5 5BG, UK\\
$^{2}$ School of Physics and Astronomy, University of Birmingham, Edgbaston, Birmingham B15 2TT, UK\\
$^{3}$ Stellar Astrophysics Centre (SAC), Department of Physics and Astronomy, Aarhus University, \\Ny Munkegade 120, DK-8000 Aarhus C, Denmark}

\date{Released 2014 Xxxxx XX}

\pagerange{\pageref{firstpage}--\pageref{lastpage}} \pubyear{2014}

\def\LaTeX{L\kern-.36em\raise.3ex\hbox{a}\kern-.15em
    T\kern-.1667em\lower.7ex\hbox{E}\kern-.125emX}

\begin{document}

\label{firstpage}

\maketitle

\begin{abstract}
The Rossiter-McLaughlin effect observed for transiting exoplanets often requires prior knowledge of the stellar projected equatorial rotational velocity (\vsini). This is usually provided by measuring the broadening of spectral lines, however this method has uncertainties as lines are also broadened by velocity fields in the stellar photosphere known as macroturbulence (\mactrb). We have estimated accurate \vsini\ values from asteroseismic analyses of main sequence stars observed by {\it Kepler}. The rotational frequency splittings of the detected solar-like oscillations of these stars are determined largely by the near-surface rotation. These estimates have been used to infer the \mactrb\ values for 28 {\it Kepler} stars. Out of this sample, 26 stars were used along with the Sun to obtain a new calibration between \mactrb, effective temperature and surface gravity. The new calibration is valid for the temperature range 5200 to 6400 K and the gravity range 4.0 to 4.6 dex. A comparison is also provided with previous \mactrb\ calibrations.  As a result of this work, \mactrb, and thus \vsini, can now be determined with confidence for stars that do not have asteroseismic data available. We present new spectroscopic \vsini\ values for the WASP planet host stars, using high resolution HARPS spectra.

\end{abstract}


\section{Introduction}
Much can be ascertained about a low-mass star from the rate at which it spins. The rotation rate of a low-mass star decreases as it evolves due to magnetized winds carrying away the angular momentum \citep{Schatzman62} and this spindown can be used to place constraints on the star's age (\citealt{Skumanich72}; \citealt{Barnes07}). The stellar rotation rate also depends on the star's mass \citep{Meibom11} and thus it is a key parameter when studying stellar evolution (\citealt{Ekstrom12}; \citealt{Gallet-Bouvier13}). 

Knowledge of the stellar rotation rate is also essential in order to understand the obliquities of transiting planets. The rotation of a star causes half of the visible disc of the star to be blueshifted as it approaches us, while the other half is redshifted as it recedes. However, a planet passing in front of the blueshifted half will weaken this radial velocity (RV) signal and temporarily cause it to be slightly redshifted, and vice versa. This is known as the Rossiter-McLaughlin (RM) effect and was first observed for binary stars by \citet{Rossiter24} and \citet{McLaughlin24}, and the technique was extended to exoplanets by \citet{Queloz00}.

The changing RV as a planet transits a star depends on the sky-projected spin-orbit angle, $\lambda$, which is the angle in the plane of the sky between the projection of the stellar spin axis and the projection of the orbital angular momentum vector of the planet. A prograde orbit ($\lambda = 0^{\circ}$) produces an antisymmetric RV signal about the transit mid-point, with the redshifted anomaly preceding the blueshifted one. A retrograde orbit ($\lambda = 180^{\circ}$) also produces an antisymmetric signal, but the redshifted anomaly follows the blueshifted one. If $0 \: \textless \: \lambda \: \textless \: 180^{\circ}$ (misaligned orbit) and the impact parameter $b \ne 0$, an asymmetric signal is produced as the planet covers more of the approaching limb than the receding one, or vice versa \citep{Gaudi-Winn07}.

Measuring planetary obliquities is important for understanding planetary evolution and migration. For instance, while the spin axis of the Sun and orbital axes of the planets in the Solar System are aligned to within $\sim$7$^{\circ}$ \citep{Beck-Giles05}, it is now known that some exoplanets have orbits that are significantly misaligned with respect to the stellar equator (e.g. \citealt{Hebrard08}; \citealt{Triaud10}; \citealt{Albrecht12}). Determining whether planets are aligned or misaligned can reveal if they migrated peacefully into their current orbits through interactions with the protoplanetary disc (e.g. \citet{Lin96}) or whether they had a complex dynamical past. Examples of processes that can lead to spin-orbit misalignment are planet-planet interactions (\citealt{Rasio-Ford96}; \citealt{Chatterjee08}; \citealt{Nagasawa08}; \citealt{Nagasawa-Ida11}; \citealt{Beauge-Nesvorny12}), the Kozai mechanism, where a planet on a close-in orbit around a star can be perturbed by a distant third body (\citealt{Kozai62}; \citealt{Fabrycky-Tremaine07}; \citealt{Morton-Johnson11}) or angular momentum variations in a hot star causing the surface of the star to rotate differently than it did when the planet was formed \citep{Rodgers12}. However, scattering and the Kozai mechanism can not be solely responsible for the distribution of hot Jupiters, and it is likely that the majority of hot Jupiters underwent a smooth migration through the protoplanetary disc, itself possibly torqued by a stellar companion (\citealt{Batygin12}; \citealt{Crida-Batygin14}).

\subsection{The importance of spectroscopic \vsini\ for the RM effect}
The measured projected rotational velocity of the star depends on the inclination angle, {\it i}, between our line of sight and the stellar rotation axis{\footnote{Not to be confused with the inclination of the planet's orbital axis to our line of sight, which can also be denoted by {\it i}, {\it I}, or $i_{\rm p}$.}}, so that the rotation measured from spectral line broadening is the projected equatorial rotational velocity \vsini. Both $\lambda$ and \vsini\ can be derived from the RM effect if {\it b} is large. However, for {\it b} $\approx$ 0, a degeneracy is introduced between $\lambda$ and \vsini\ so that a prior knowledge of \vsini\ is essential in order to extract $\lambda$ (see \citet{Albrecht11} for more details). The \vsini\ prior is usually obtained from the rotational broadening of stellar spectral lines and as the \vsini\ is just one of many factors that broadens spectral lines, it is difficult to extract an accurate value.

A significant contribution to spectral line broadening comes from velocity fields in the stellar photosphere. In 1D model atmospheres, these velocity fields are represented by microturbulence (\mictrb) and macroturbulence (\mactrb), which unfortunately have misleading names as they physically have very little to do with turbulence. The size of the microturbulent ``cell'' is defined as being less than the mean free path of the photon, whereas \mactrb\ represents velocities that occur where the cell is larger than the unit optical depth. The \mictrb\ changes the equivalent width (EW) of spectral lines, however \mactrb\ does not change the EW (\citealt{Magain84}; \citealt{Mucciarelli11}).

If individual spectra from each macro cell could be obtained, they would show a radial velocity shift. However, as many are viewed simultaneously because the stellar disc is unresolved, they become averaged and the overall effect is to broaden the line profile. Macroturbulence can be modelled assuming that both radial and tangential motion will take place, which changes the line profile shape to give broadened wings and a ``cusp'' shaped core (see \citet{Gray08} for more details). Unfortunately, disentangling the rotational profile from the radial-tangential \mactrb\ profile is difficult, leading to a degeneracy between the two. While they can be disentangled using Fourier techniques \citep{Gray84}, this method requires spectra of extremely high resolution and a high signal-to-noise ratio (S/N), and such spectra are not always available.

Therefore, when a spectroscopic \vsini\ is required for studying the RM effect, the \mactrb\ must be known first. Relationships between stellar effective temperature (\teff) and \mactrb\ exist (e.g. \citealt{Gray84}; \citealt{Saar-Osten97}; \citealt{Gonzalez98}; \citealt{Bruntt10a} and the upper limit \mactrb\ relationship of \citealt{Valenti-Fischer05}), however the choice of \mactrb\ will influence the \vsini, creating additional uncertainties. If the \vsini\ is determined from a reliable, external method, such as via asteroseismology, then it is possible to break this degeneracy. 

A reliable determination of \vsini\ can also be used to estimate the stellar inclination along the line of sight if the \vsini\ is combined with the rotation period, which can be determined from variations in the light curve due to starspots, and stellar radius (\citealt{Hirano12}; \citealt{Hirano14}). Once {\it i} is known, it is also possible to determine the true spin-orbit angle, $\psi$, between the stellar spin axis and the orbit of the planet if $\lambda$ and the orbital inclination of the planet are known from the transit.

Thus, the motivation of this work was to determine the \mactrb\ from the spectra of 28 stars in \citet{Bruntt12} by fixing the \vsini\ to the asteroseismic value. Section~\ref{method} details how asteroseismology was used to measure the rotational splitting of modes, and thus the \vsini. The method of fitting the \mactrb\ to the spectral lines is also outlined in this section. The resulting calibration between \mactrb, \teff, and surface gravity (\logg) as obtained using the Sun and 26 out of the 28 stars is given in Section~\ref{results}. Section~\ref{discussion} compares this work to other calibrations that are frequently used in the literature, and gives the new spectroscopic \vsini\ values for the WASP stars and we conclude in Section~\ref{conclusion}.

\section{Method}
\label{method}
\subsection{Principles of asteroseismic rotation}
The characteristics of stars are written within their pulsation frequencies. NASA's {\it Kepler} mission \citep{Borucki10} has revealed solar-like oscillations in hundreds of main sequence stars \citep{Chaplin14} and thousands of red giants \citep{Hekker11}. These oscillations can be used to determine stellar properties, such as mean density, surface gravity, age, mass and radius ($R_{s}$).

Solar-like oscillations are standing waves that are excited by turbulent convection near the surface of the star. There are two types of standing waves associated with solar-like oscillations. Acoustic oscillations, known as p-modes, rely on the gradient of the pressure as a restoring force, where as g-modes use gravity as a restoring force. In main sequence stars, g-modes are confined beneath the convection zone which means that these perturbations are extremely weak at the surface and thus are difficult to detect. However, for evolved stars, mixed modes can be detected as the increased core density causes the g-mode frequencies from the stellar interior to increase to the point that they are comparable to p-mode frequencies. The coupling between the p-mode and g-mode cavities causes the modes to exchange nature. Therefore, these mixed modes behave as p-modes in the envelope and g-modes in the interior.

Each oscillation mode in a star is described by a characteristic frequency for particular values of {\it n}, {\it l} and {\it m}. The overtone of the mode, {\it n}, is the number of radial nodes, or nodal shells. The degree of the mode, {\it l}, is the number of nodal lines at the surface and the azimuthal order of the mode, {\it m}, is the number of surface nodes that cross the equator and is given as (2{\it l} + 1). In the absence of rotation, the {\it m} frequencies will all be the same and the mode frequency is given as $\nu_{nl}$. For a rotating star, this degeneracy is lifted so that the non-radial mode is now split into a multiplet of the {\it m} components and the frequency splitting is given as $\delta \nu_{nlm}$. For solar-like rates of rotation we can ignore the effects of fictitious forces, i.e. the Coriolis force ($\approx$ 1 per cent) and centrifugal distortion (\citealt{Reese06}, \citealt{Baglin10}). The frequency of the mode is thus
\begin{equation}
\nu_{nlm} \equiv \nu_{nl} + \delta \nu_{nlm},
\end{equation}
with
\begin{equation}
\delta \nu_{nlm} \approx \frac{m}{2\pi}\int_{0}^{R}\int_{0}^{\pi} K_{nlm}(r, \theta) \Omega(r,\theta)r \; {\rm d} r \; {\rm d} \theta,
\end{equation}
where $\Omega (r, \theta)$ is the radially and co-latitudinally dependent internal angular velocity, and $K_{nlm}(r, \theta)$ is a weighting kernel that reflects the sensitivity of the mode to the internal rotation.

Here we apply a treatment that assumes solid body rotation and symmetric frequency splitting \citep{Ledoux51}. Solid body rotation clearly ignores the effects of latitudinal differential rotation and radial differential rotation while symmetric frequency splitting neglects any contribution from near surface magnetic fields \citep{Dziembowski-Goode97}.  These effects are expected to be small for the observable high-$n$, low-$l$ p modes.  In fact, in Sun-as-a-star data with very much higher signal-to-noise ratios, these effects are difficult to observe and should not pose a problem here \citep{Chaplin11c}.  We can then simplify by setting $\delta \nu_{nlm} = m \left< \delta \nu_{s} \right> $, i.e. all frequency splitting has a common value.  This simplification reduces the complexity of the parameter estimation while also providing better parameter constraint.

As non-radial modes are not spherically symmetric, the inclination will be a factor in detecting the frequency splitting. For example, in an ideal situation, an {\it l} = 1 mode has {\it m} = $\pm$ 1 components that can be easily discerned at {\it i} = 90$^{\circ}$. For this scenario, the {\it m} = 0 component does not contribute to the line profile at all because at {\it i} = 90 the intensity perturbations from the northern and southern hemispheres cancel out. For lower values of inclination, the splitting evident from the {\it m} = $\pm$ 1 modes will become less evident as blending becomes more prominent. At {\it i} = 0, the contribution from the {\it m} = $\pm$ 1 components vanishes completely, leaving only the {\it m} = 0 component, which presents itself as a single, unsplit line profile \citep{Chaplin13}.

When the contributions to the observed stellar intensity across the visible disc depend only on the angular distance from the disc centre, as is the case for photometric observations, and there is energy equipartition between the different $m$ components, as is the case for the modest rates of rotation considered here, we may use the formulation of \citet{Gizon-Solanki03}.
For this formulation the disc integrated amplitudes of the $m$ components depend only on the observers angle of inclination to the pulsation axis, {\it i}. This dependence (in power) may be written as
\begin{equation}
\xi_{l}^{|m|}(i) = \frac{(l - |m|)!}{(1 + |m|)!}\left[ P_{l}^{|m|} (\cos i)\right]^{2},
\end{equation}  
where $P_{l}^{|m|}$ is the Legendre function and the sum of $\xi_{l}^{|m|}(i)$ over $m$ is normalized to unity.  Hence, measuring the relative power of the azimuthal components provides a direct estimate of the pulsational angle of inclination.  In the absence of a very strong magnetic field one expects the rotational and pulsational axes to be aligned and hence we have a measure on the rotational angle of inclination.  One should note the inherent symmetries in the $m = |m|$ components mean we cannot discriminate between $i$ and $- i$, and $\pi - i$ and $\pi + i$. 

The linewidth of the mode, $\Gamma$, will have an affect on the ability to detect frequency splitting. As \teff\ increases, $\Gamma$ will also increase \citep{Appourchaux12}, which makes \nusini\ harder to determine for hotter stars because the blending becomes more prominent. For instance if the linewidth or the splitting is similar to the small frequency separation, it can make it difficult to distinguish between {\it l} = 0 and {\it l} = 2 modes, thus hampering the ability to fit a unique solution \citep{Barban09}.

\subsection{Selection of stars}
We have selected stars that fit our criteria for both spectroscopic and asteroseismic analyses. High resolution spectra were required in order to distinguish between different types of broadening in the spectral lines. The spectra used for this work were obtained by \citet{Bruntt12} using the ESPaDOnS spectrograph at the 3.6\,m Canada-France-Hawaii Telescope, and the Narval spectrograph at the 2\,m Bernard Lyot Telescope. ESPaDOnS has a resolving power (R) of $\sim$80\,000 (\citealt{Donati04}; \citealt{Donati06}) and Narval, which is an almost identical spectrograph, has a resolving power of $\sim$75\,000. 

For the results of an asteroseismic analysis to be comparable to spectroscopic measures of rotation, we require stars where the peak sensitivity of the modes of oscillation is located close to the stellar surface, approximately $> 0.95 R_{s}$.  For this to be true, we select only stars that do not show evidence of modes of mixed character that show greater sensitivity to the deep stellar interior (see \citet{Deheuvels12} and \citet{Deheuvels14} for more details on mixed modes in the context of rotation). Modes of mixed character can be easily identified by inspection of the asteroseismic echelle diagram (for examples of categorisation see \citet{Appourchaux12b}) and hence rejected from our sample.  In addition, stars with low S/N for individual modes do not provide sufficient constraint for asteroseismic rotation to be estimated.  Hence, stars showing fewer than five radial orders of individual frequencies were rejected.  

Applying the spectroscopic and asteroseismic constraints gives the 35 targets listed in Table~\ref{mactresults}.

\subsection{Determination of rotation}
We used {\it Kepler} short-cadence observations \citep{Gilliland10} taken from Q5 - Q11 which were generated using simple aperture photometry \citep{Jenkins10} corrected for instrumental effects following the methods described by \cite{Garcia11}. The desired rotation estimates are an output of ``peak-bagging'' \citep{Appourchaux03}, which is modelling of the observed power spectrum.  The model applied can be decomposed into two categories of phenomenon, background (noise to seismologists) and modes of oscillation (signal to seismologists).

The background $B(\nu)$ is modelled using a Harvey component \citep{Harvey85} in addition to a white component to account for photon shot noise.  This gives
\begin{equation}
B(\nu) = W + \frac{4 \sigma_{k}^{2} \tau_{k}}{1 + (2\pi \tau_{k} \nu)^{c}},
\end{equation}   
where $\sigma_{k}$ is related to the rms amplitude of the signal, $\tau_{k}$ is the characteristic time scale of the decaying autocorrelation function, and the exponent $c$ is related to the shape of the excitation of the function.

The oscillations $O(\nu)$ are modelled as a sum of Lorentzian profiles that characterize the frequency-power limit spectrum of stochastically excited and intrinsically damped modes.  The model is then
\begin{equation}
O(\nu) = \sum_{n^{'},l} \sum_{m=-l}^{l} \frac{\xi_{l}^{|m|}(i_{s}) H_{n^{'}l}}{1 + \left(2/\Gamma_{n^{'}}\right)^{2} \left(\nu - \nu_{n^{'}l} - m \left< \delta \nu_{s} \right> \right)^{2}},
\end{equation} 
where $H_{n^{'}l}$ is the mode height and $n^{'}$ is the dummy variable such that $n^{'} = n$ for $l=0,1$ and $n^{'} = n-1$ for $l=2,3$.  This defines the basic model we use but we apply a number of accepted simplifications to reduce the size of the parameter space to a tractable problem.

First, we assume that the mode linewidth across a certain $n^{'}$ does not vary.  That is we use one value of linewidth for groups of $l=2,0,3,1$ modes.  Retrospectively, we find that this assumption is sensible to well within the uncertainty of the linewidth parameters returned.

Secondly, mode heights are constrained by the relation $H_{n^{'},l} = H_{n^{'},0} V_{l}^{2}$, where $V_{l}^{2}$ is commonly referred to as the degree visibility.  The degree visibilities can be estimated theoretically (\citealt{Bedding96}; \citealt{Ballot11b}) but here we leave the values as free parameters to be determined during the peak bagging.

Finally, and crucially for this analysis, we make two changes of variable to reduce the correlations in the parameter space.  In the first instance we do not explore the highly anti-correlated parameter space defined by the mode height and line width.  Instead we change variables to explore the much less correlated mode amplitude squared ($A^{2}$) and line width space.  This transformation is straightforwardly applied given that
\begin{equation}
A^{2} = \frac{2}{\pi} H \Gamma.
\end{equation}
Equally as simple is the transformation of variables for rotation that means we explore the parameter space of the angle and $\delta \nu_{s} \sin {i_{s}}$, the so called ``projected splitting''.  These two parameters are only lightly correlated, something that is essential for the robust marginalisation of each posterior probability density. By using \nusini\ along with an accurate determination of stellar radius from asteroseismology \citep{Chaplin14}, the \vsini\ of the star can be determined via \citep{Chaplin13}
\begin{equation}
\vsini \equiv 2 \pi R_{s} \, \nusini.
\end{equation}

Parameter determination for the model given the observations was performed using Markov Chain Monte Carlo (MCMC) methods (for examples see \cite{Benomar09b, Handberg-Campante11}).  We applied uniform priors to the angle of inclination between $0^{\circ}$ and $90^{\circ}$ and the projected splitting between $0$ and $5.0 \rm \; \mu Hz$.

\subsection{Determination of macroturbulence}
\label{mact-determination}
We selected a set of lines from \citet{Doyle13} with which to fit the \mactrb, and these are listed in Table~\ref{mact-lines}, including the excitation potential ($\chi$) and oscillator strength (\loggf). These lines are as unblended as possible over a large \teff\ range. However, in cooler stars blending will become an issue for some of these lines, such as Ti~{\sc ii} 5418 \AA, in which case they are rejected by visual inspection. Stars that are slightly metal-poor will also have fewer measurable lines. In stars with a relatively high \vsini\ ($\gtrsim$ 12 \kms), blending also becomes an issue as the selected lines will become broadened and encroach on other nearby lines. Unresolved blends were rejected in these stars, however due to the lack of suitable lines in high \vsini\ stars, resolved blends were still used. 

\begin{table}
\centering
\caption{Spectral lines used to fit macroturbulence}
\begin{tabular}{cccc} \hline
Element & Wavelength (\AA) & $\chi$ (eV) & \loggf\ \\ \hline
Cr~{\sc i}  & 5238.964 & 2.709 & $-$1.305 \\
Ti~{\sc i}  & 5295.780 & 1.067 & $-$1.633 \\
Y~{\sc ii}  & 5402.774 & 1.839 & $-$0.510 \\
Ti~{\sc ii} & 5418.751 & 1.582 & $-$2.110 \\
Fe~{\sc ii} & 5425.257 & 3.199 & $-$3.220 \\
Fe~{\sc i}  & 5538.517 & 4.218 & $-$3.244 \\
Fe~{\sc i}  & 5576.090 & 3.430 & $-$1.000 \\
Fe~{\sc i}  & 5651.470 & 4.473 & $-$2.000 \\
Ca~{\sc i}  & 5867.563 & 2.933 & $-$1.570 \\
Ni~{\sc i}  & 6111.066 & 4.088 & $-$0.870 \\
Fe~{\sc i}  & 6151.617 & 2.176 & $-$3.299 \\
Fe~{\sc i}  & 6200.319 & 2.609 & $-$2.437 \\
Ni~{\sc i}  & 6204.600 & 4.088 & $-$1.100 \\
Ni~{\sc i}  & 6223.980 & 4.105 & $-$0.910 \\
Fe~{\sc i}  & 6252.554 & 2.404 & $-$1.687 \\
Ti~{\sc i}  & 6258.104 & 1.443 & $-$0.355 \\
Ni~{\sc i}  & 6378.247 & 4.154 & $-$0.830 \\
Ni~{\sc i}  & 6772.313 & 3.658 & $-$0.980 \\
Fe~{\sc i}  & 6810.257 & 4.607 & $-$0.986 \\
Fe~{\sc i}  & 6857.249 & 4.076 & $-$2.150 \\

\hline
\end{tabular}
\label{mact-lines}
\end{table}

We used the software {\sc uclsyn} (\citealt {Smith92}; \citealt {Smalley01}) to perform the analyses. {\sevensize\ ATLAS}9 models atmospheres without convective overshooting are used \citep{Castelli97} and local thermodynamic equilibrium is assumed.

We first determined the  radial-tangential \mactrb\ for the Kitt Peak Solar Atlas \citep{Kurucz84}. The \vsini\ was fixed to  1.9 \kms\ \citep{Gray77}, and \mictrb\ was assumed to be 1.0 \kms. We checked all lines individually by eye to ensure that the fitting was correct and to eliminate any lines with bad fits. The final \mactrb\ value given is the average of all the lines used. The \mactrb\ determined for the Kitt Peak Solar Atlas is 3.21 $\pm$ 0.27 \kms, and as the resolution of 300\,000 means that the instrumental broadening is negligible in this spectrum, this \mactrb\ value was deemed to be the solar value for the purpose of this paper. This is reasonably consistent with the range of values (3.1 \kms\ for strong lines and 3.8 \kms\ for weak lines) determined from the Fourier analysis of \citet{Gray77}.

In order to truly disentangle the rotational and macroturbulent broadening in the line profiles, it is imperative to first know how the spectrograph itself broadens the lines. The telluric lines at $\sim$6880 \AA\ in the ESPaDOnS and Narval solar spectra suggest a resolution between $\sim$75\,000 and $\sim$81\,000, however an exact value can not be determined as the same resolution value does not fit all lines.

Using the nominal resolution of 81\,000 for ESPaDOnS gave \mactrb\ = 3.48 $\pm$ 0.40 \kms. Similarly, the nominal resolution of 75\,000 for Narval gave \mactrb\ = 3.04 $\pm$ 0.50 \kms. Although these values agree with the Kitt Peak Solar Atlas within the errors, the discrepancies can be ascribed to a slightly incorrect value of the spectral resolution adopted for ESPaDOnS and Narval. As a check on the resolution values, we varied the the average resolution of the synthetic spectra until the \mactrb\ equal to the Kitt Peak Solar Atlas value. The HARPS solar spectrum \citep{Dall06} was also included. The final resolution and \mactrb\ for each solar spectrum are given in Table~\ref{solar}.

\begin{table}
\centering
\caption{Resolution and \mactrb\ for each solar spectrum}
\begin{tabular}{lll} \hline
Spectrograph & Resolution & \mactrb\ (\kms)  \\ \hline
Solar Atlas  & 300\,000 & 3.21 $\pm$ 0.27 \\
HARPS (day sky)	     & 98\,000  & 3.21 $\pm$ 0.19 \\
ESPaDOnS (twilight sky)    & 76\,000  & 3.21 $\pm$ 0.53 \\
Narval (Moon)	     & 80\,000  & 3.20 $\pm$ 0.49 \\

\hline
\end{tabular}
\label{solar}
\end{table}

For the {\it Kepler} stars, we used the \teff, \logg\ and \mictrb\  from \citet{Bruntt12}, where \logg\ was determined from asteroseismology. The errors are given as 60 K, 0.03 dex and 0.05 \kms\ respectively. Including the resolution determined from the solar spectra and fixing the \vsini\ to the asteroseismic value, the radial-tangential \mactrb\ was determined for each star.

\begin{figure}
\centering
\includegraphics[height=\columnwidth,angle=270]{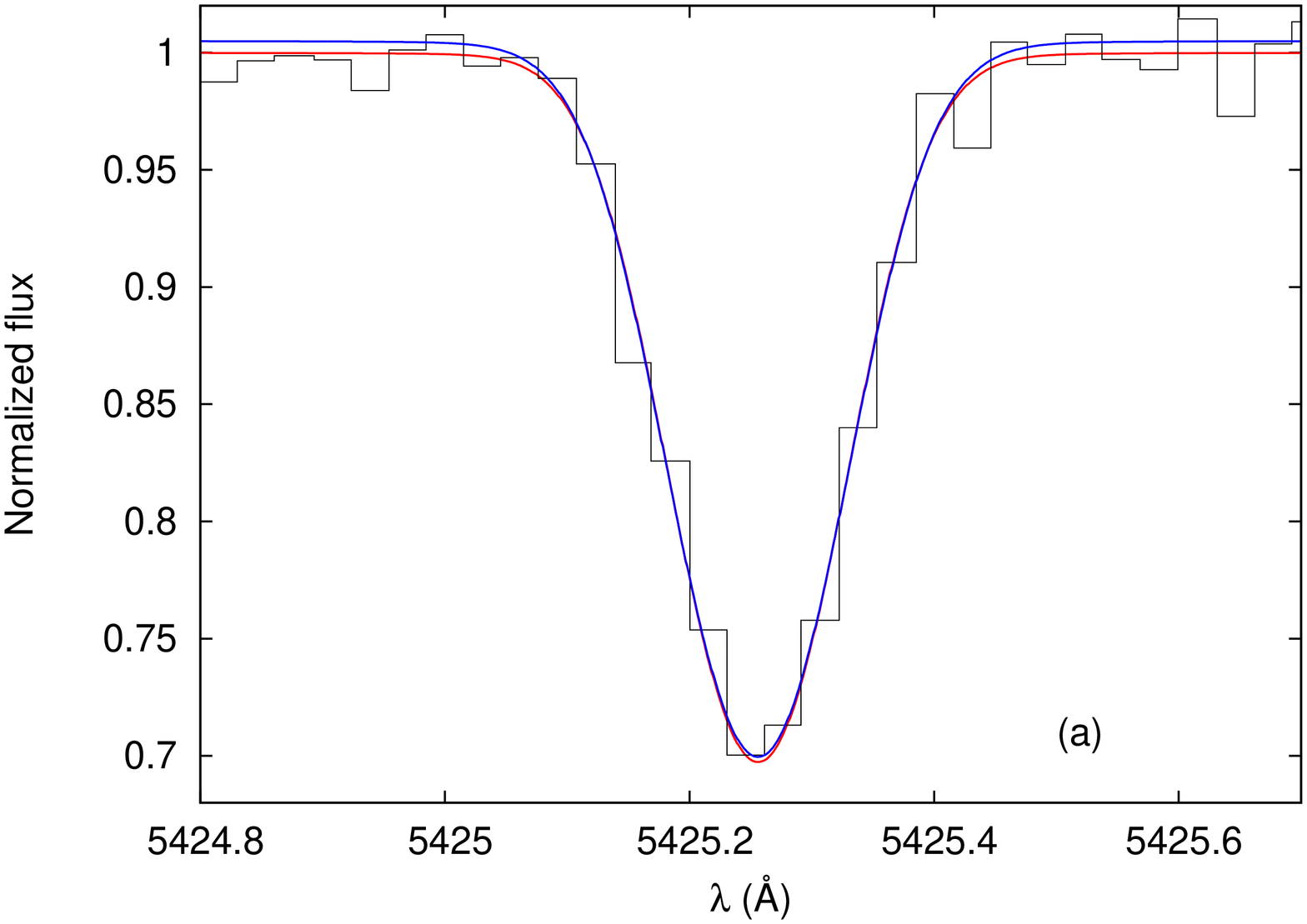}
\includegraphics[height=\columnwidth,angle=270]{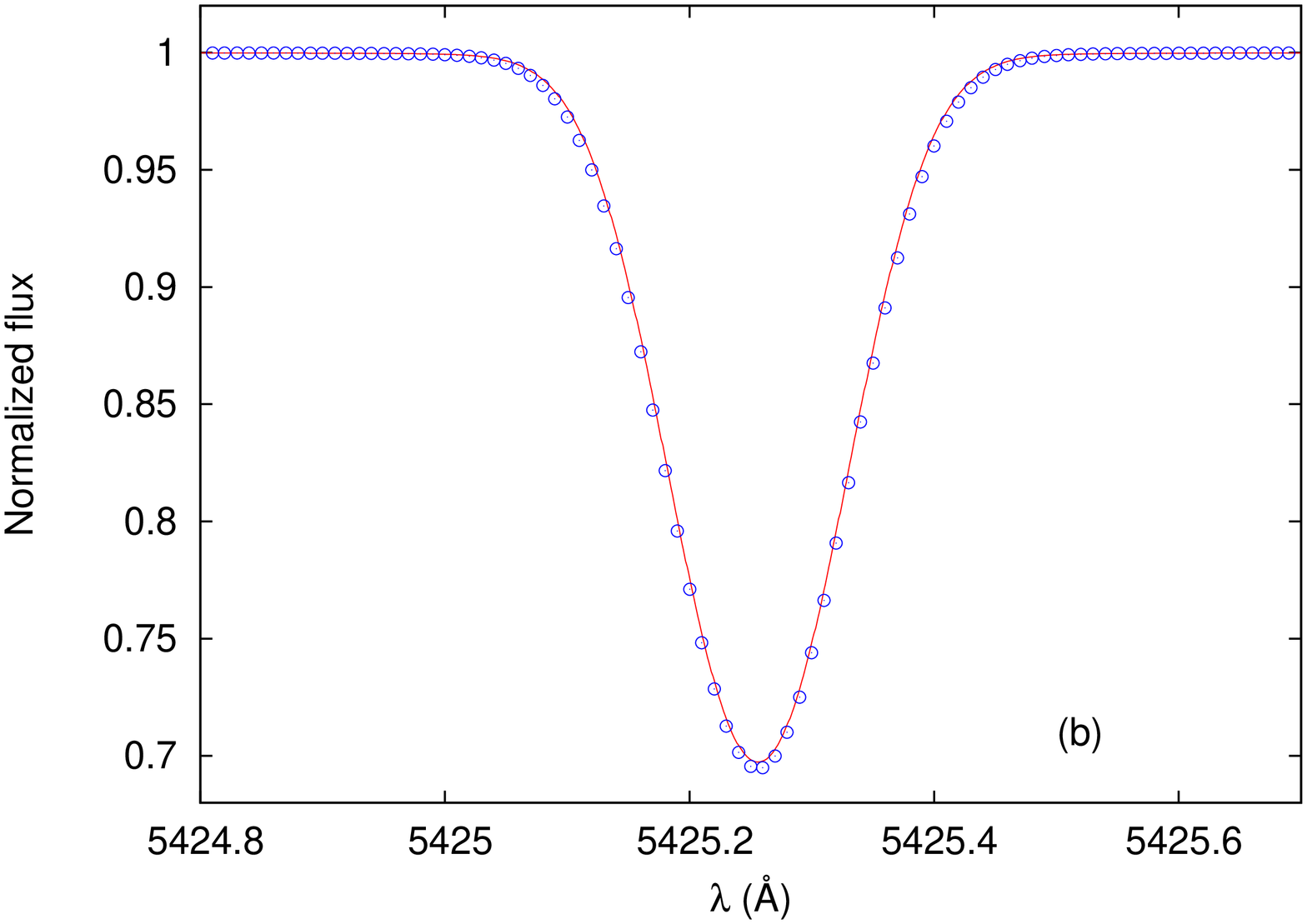}
\caption{(a) The S/N of $\sim$100:1 makes continuum placement ambiguous, as indicated by two different synthetic lines (in red and blue) which differ by 0.5 per cent at the continuum level for a Fe~{\sc ii} line in KIC 8228742. (b) The two synthetic lines shown are normalized to 1 and show a subtle difference in the \mactrb\ fit for the same spectral line, which leads to two different \mactrb\ values of 3.34 (solid line) and 3.76 (circles) \kms.}
\label{norm-comp}
\end{figure}

The main contribution to the \mactrb\ error for each star is from the scatter in the \mactrb\ values for the different lines. This is because the spectra have a S/N of 100:1 or less at 5500 \AA, which means that the noise at the continuum level makes it difficult to determine the exact position of the continuum as seen in Fig.~\ref{norm-comp} (a). Adjusting the continuum by even 0.5 per cent can result in a different \mactrb\ value of up to 0.5 \kms. Fig.~\ref{norm-comp} (b) shows the two different synthetic lines of (a) normalized to 1. Upon close inspection, the two are not in exact agreement and they return two different \mactrb\ values of 3.34 and 3.76 \kms.

The \vsini\ error is included in the \mactrb\ error, however it has a negligible impact except for KIC 10355856 where the large \vsini\ error propagates into the \mactrb\ error. 

Different spectral analysis techniques can yield different values of \mictrb\ for the same spectrum, and the choice of \mictrb\ can affect the resulting \mactrb. For instance, using a \mictrb\ of 1.00 and 0.85 \kms\ in the Kitt Peak solar atlas yields a \mactrb\ of 3.21 $\pm$ 0.27 \kms\ and 3.33 $\pm$ 0.28 \kms\ respectively.  Thus, the choice of \mictrb\ can have a subtle influence on \mactrb. In order to account for this, a \mictrb\ error of 0.15 \kms\ was added in quadrature to the \mactrb\ error.

The instrumental broadening needs to be input before the \mactrb\ can be determined, however using the wrong resolution will introduce an additional error to the \mactrb. Increasing/decreasing the resolution of the ESPaDOnS and Narval solar spectra by 5000 increases/decreases the \mactrb\ by 0.19 and 0.14 \kms\ respectively and this was accounted for in the \mactrb\ errors.

\section{Results}
\label{results}
The asteroseismic \vsini\ and the derived \mactrb\ are listed in Table~\ref{mactresults}, and stars are identified with their {\it Kepler} Input Catalog (KIC) numbers. Out of 35 stars with asteroseismic \vsini\ values available, we were able to measure the \mactrb\ for 28 stars, and 26 of these were used for the calibration, along with the Sun. Stars that were not included in the calibration do not have a \mactrb\ value listed in Table~\ref{mactresults}.

The \mactrb\ is plotted against \teff\ in Fig.~\ref{calibration}, showing a clear increase in \mactrb\ with increasing \teff. The plot also shows that there appears to be some dependency on \logg\ among the dwarfs, indicating that \logg\ should be accounted for even within the same luminosity class. There are two young red giants and one subgiant from \citet{Deheuvels14} also included on the plot, but not included in the calibration. These show that \mactrb\ is higher for the giants as expected \citep{Gray08}, although it should be noted that the \nusini\ values for these stars are upper limits.  The fit to the data is expressed as a function of \teff\ and \logg\ via

\begin{multline}
\mactrb\ = 3.21 + 2.33 \times 10^{-3}(\teff-5777) \\ 
 + 2.00 \times 10^{-6}(\teff - 5777)^{2} - 2.00(\logg - 4.44).
\label{fit}
\end{multline}

\noindent
The zero points were set to yield the \mactrb\ value for the Sun. This calibration is valid for the \teff\ range 5200 to 6400 K, and the \logg\ range 4.0 to 4.6 dex. {A total error of 0.73 \kms\ is determined from adding in quadrature the the rms scatter of the fit (0.37 \kms) and the mean of the \mactrb\ errors with a 3-$\sigma$ clipping rejection criterion (0.62 \kms).
\begin{table*}
\centering
\begin{minipage}{120mm}
\caption{The asteroseismic \vsini\ and \mactrb\ as determined in this work. The asteroseismic \vsini\ is too high to fit the line profiles in some stars, in which case no \mactrb\ is given. The \teff, \logg\ and \mictrb\ are from \citet{Bruntt12}, with errors of 60 K, 0.03 dex and 0.05 \kms\ respectively.}
\begin{tabular}{p{1cm} p{1cm} p{1cm} p{0.6cm} p{0.5cm} p{0.8cm} p{1.6cm} p{1.6cm}} \hline
KIC & HD & HIP &\teff\ & \logg\ & \mictrb\ & $\vsini_{\rm astero}$ & \mactrb\  \\ 
    & & & K & & \kms\ & \kms\ & \kms \\ \hline
1435467 &  & & 6264 & 4.09 & 1.45 & 10.58 $\pm$ 0.70 & 5.61 $\pm$ 1.35 \\ 
2837475	&  179260 & & 6700 & 4.16 & 2.35 & 21.50 $\pm$ 0.96 & 12.07 $\pm$ 2.31 \\
3427720 &  & & 6040 & 4.38 & 1.16 & 1.07 $\pm$ 0.63 & 3.80 $\pm$ 0.82 \\
3456181 &  & & 6270 & 3.93 &	1.53 & 10.89 $\pm$ 0.52 & \\
3632418	& 179070 & 94112 & 6190 & 4.00 & 1.42 & 7.75 $\pm$ 0.46 & 4.87 $\pm$ 0.87 \\
3656476 &  &  & 5710 & 4.23 & 1.02 & 1.13 $\pm$ 0.18 & 3.56 $\pm$ 0.49 \\	
4914923	&  & 94734 & 5905 & 4.21 & 1.19 & 2.46 $\pm$ 0.39 & 4.13 $\pm$ 0.42 \\
5184732 &  & & 5840 & 4.26 & 1.13 & 3.11 $\pm$ 0.19 & 3.67 $\pm$ 0.64 \\
6106415	& 177153 & 93427 & 5990 & 4.31 & 1.15 & 3.66 $\pm$ 0.14 & 4.14 $\pm$ 0.59 \\
6116048 &  & & 5935 & 4.28 & 1.02 & 3.47 $\pm$ 0.16 & 4.02 $\pm$ 0.57 \\ 
6225718 & 187637 & 97527 & 6230 & 4.32 & 1.38 & 15.46 $\pm$ 1.13 & \\
6508366	&  & & 6354 & 3.94 & 1.52 & 20.59 $\pm$ 0.95 & 9.83 $\pm$ 1.35 \\
6679371 &  & & 6260 & 3.92 & 1.62 & 18.53 $\pm$ 0.92 & \\
6933899 &  & & 5860 & 4.09 & 1.15 & 1.99 $\pm$ 0.30 & 4.19 $\pm$ 0.63 \\
7103006 &  & & 6394 & 4.01 & 1.58 & 13.46 $\pm$ 1.04 & \\
7206837 &  & & 6304 & 4.17 & 1.29 & 7.82 $\pm$ 1.06 & \\
7680114	&  & & 5855 & 4.18 & 1.10 & 2.49 $\pm$ 0.27 & 3.65 $\pm$ 0.53 \\
7871531 &  & & 5400 & 4.49 &	0.71 & 1.22 $\pm$ 0.27 & 2.81 $\pm$ 0.52 \\
7940546 & 175226 & 92615 & 6264 & 3.99 & 1.56 & 9.17 $\pm$ 0.42 & \\
7970740 & 183606 & & 5290 & 4.58 &	0.68 & 0.70 $\pm$ 0.20 & 2.50 $\pm$ 0.74 \\
8006161	&  & 91949 & 5390 & 4.49 & 1.07 & 1.20 $\pm$ 0.08 & 2.22 $\pm$ 0.58 \\
8228742 &  & 95098 & 6042 & 4.02 & 1.30 & 5.15 $\pm$ 0.59 & 4.22 $\pm$ 0.85 \\
8394589	&  & & 6114 & 4.32 & 1.23 & 4.92 $\pm$ 0.33 & 5.09 $\pm$ 0.65 \\
8694723 &  & & 6120 & 4.10 & 1.39 & 4.19 $\pm$ 0.78 & 6.28 $\pm$ 1.27 \\
9098294	&  & & 5840 & 4.30 & 1.01 & 2.11 $\pm$ 0.36 & 3.71 $\pm$ 0.69 \\
9139151	&  & 92961 & 6125 & 4.38 & 1.22 & 4.75 $\pm$ 0.31 & 3.98 $\pm$ 0.74 \\
9139163 & 176071 & 92962 & 6400 & 4.18 &	1.31 & 10.15 $\pm$ 0.81 & \\
9812850 &  & & 6325 & 4.05 & 1.61 & 12.04 $\pm$ 0.96 & 6.06 $\pm$ 1.27\\
9955598	&  & & 5410 & 4.48 & 0.87 & 1.29 $\pm$ 0.12 & 2.51 $\pm$ 0.76 \\
10355856 &  & & 6350 & 4.08 & 1.55 & 5.74 $\pm$ 2.72 & 5.75 $\pm$ 2.56 \\
10454113 &  & 92983 & 6120 & 4.31 & 1.21 & 3.83 $\pm$ 0.51 & 4.81 $\pm$ 0.64 \\
10644253 &  & & 6030 & 4.40 & 1.14 & 0.62 $\pm$ 0.81 & 3.85 $\pm$ 0.56 \\
10963065 &  & & 6060 & 4.29 & 1.06 & 3.61 $\pm$ 0.25 & 4.35 $\pm$ 0.46 \\
11244118 &  & & 5745 & 4.09 & 1.16 & 1.67 $\pm$ 0.22 & 3.66 $\pm$ 0.73 \\
12009504 &  & & 6065 & 4.21 & 1.13 & 7.36 $\pm$ 0.37 & 4.41 $\pm$ 0.60\\

\hline

\end{tabular}
\label{mactresults}
\end{minipage}
\end{table*}    

\begin{figure*}
\centering
\begin{minipage}{170mm}
\includegraphics[height=\columnwidth,angle=270]{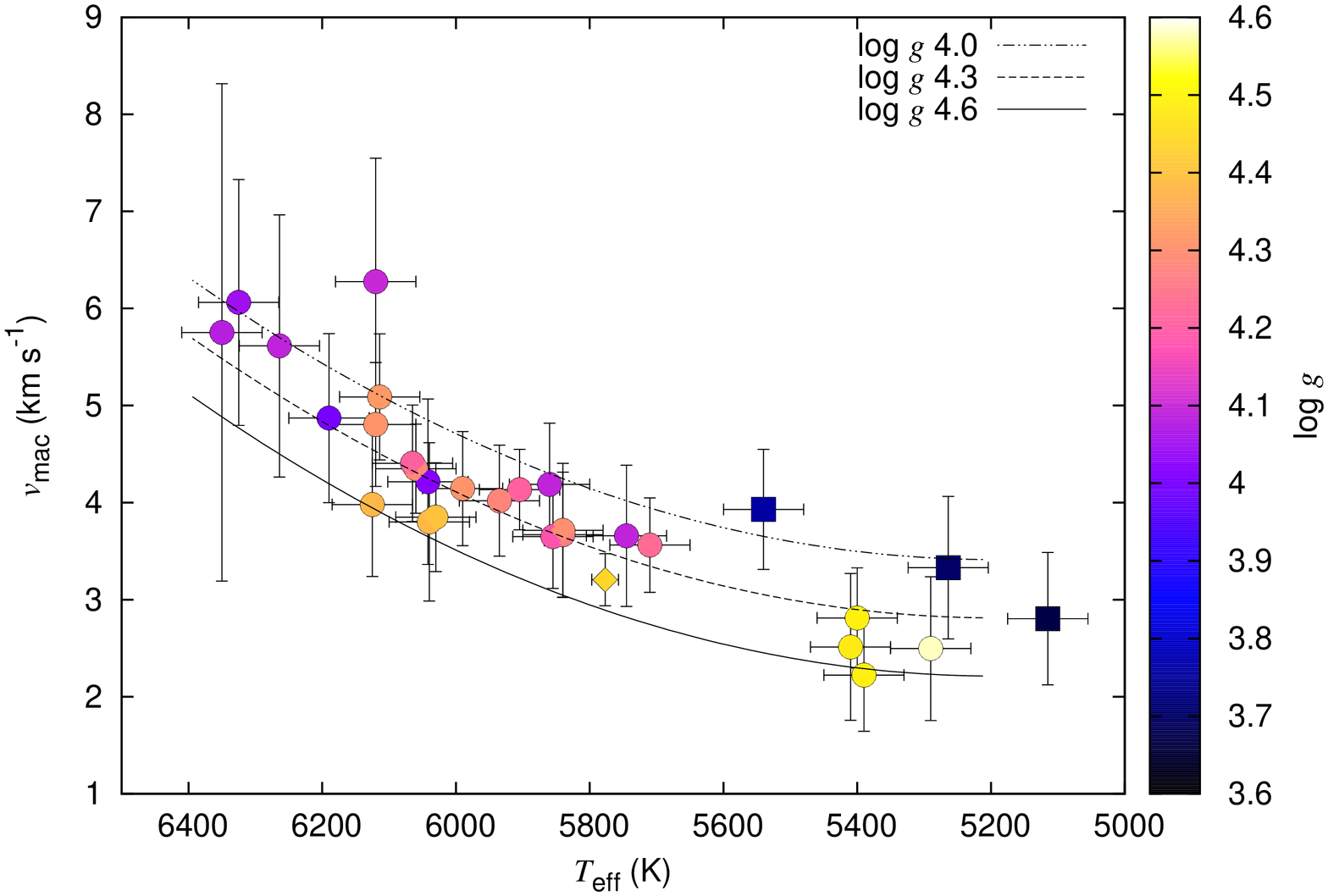}
\caption[]{Macroturbulence is seen to increase with increasing \teff, however there also seems to be some \logg\ dependence. The circles represent the stars used in this study, the diamond represents the Sun, and the squares are the red giants from \citet{Deheuvels14}. The red giants are not included in the calibration. }
\label{calibration}
\end{minipage}
\end{figure*}

There are two outliers, KIC 2837475 and KIC 6508366, not shown on the plot which have abnormally high \mactrb. Judging by the extent that the wings of the lines are broadened, it is possible that this effect could be real. However, both stars also have high \vsini\ ($\sim$20 \kms), meaning that only a few lines are available to fit \mactrb\ and it is quite difficult to obtain a reasonable fit. KIC 2837475 has a large discrepancy between the spectroscopic and asteroseismic \logg\ with the former being  0.35 dex higher than the latter.  \citet{Bruntt12} finds that the spectroscopic \logg\ is on average 0.08 $\pm$ 0.07 dex higher than the asteroseismic values, but cannot explain the discrepancy of KIC 2837475.

Seven of the stars have an asteroseismic \vsini\ that is clearly too high to fit the spectra as it does not allow for any \mactrb\ broadening. This means that the wings of the lines cannot be fit and the cores of the synthetic lines are too shallow, as shown in Fig.~\ref{highvsini}. This implies that another mechanism is changing the shape of the line profiles. Latitudinal differential rotation, low inclination and limb darkening have the effect of narrowing line profiles \citep{Reiners-Schmitt02}, which could explain why the \vsini\ will not fit. However, limb darkening would not explain why this effect is only seen in some stars and not others with similar parameters. As \nusini\ cannot be well constrained for stars with a low inclination, the discrepancy in the line profiles cannot be caused by low inclination. While there is some correlation between {\it i} and \nusini, it is still possible to place lower limits on the inclination, which show that $i \gtrsim$ 40 for these stars and most have $i \gtrsim$ 60. Therefore, the most likely explanation for the shapes of these line profiles is differential rotation. This will be discussed further in a forthcoming paper.

\begin{figure}
\centering
\includegraphics[height=\columnwidth,angle=270]{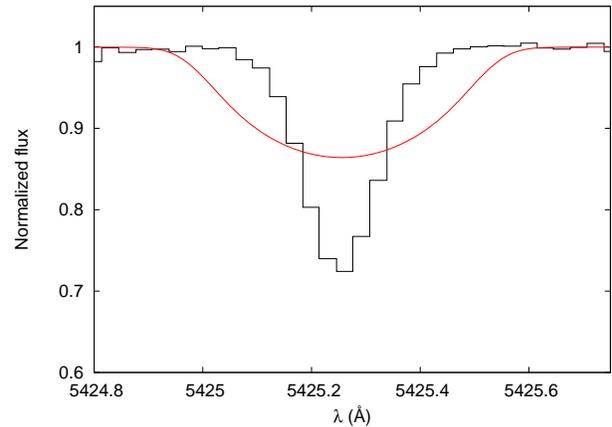}
\caption[]{The synthetic spectrum clearly does not fit the observed spectrum of KIC 6225718. The synthetic line, in red, has $\vsini_{\rm astero}$ = 15.46 $\pm$ 1.13 \kms\ and \mactrb\ = 0 \kms.}
\label{highvsini}
\end{figure}

\section{Discussion}
\label{discussion}
\subsection{Comparison with previous calibrations}
\label{comparisons}
In this section we compare our calibration with some of the most widely used \mactrb\ relationships in the literature; namely those of \citet{Gray84}, \citet{Valenti-Fischer05} and \citet{Bruntt10a}. Fig.~\ref{compare} shows the different calibrations on the \mactrb--\teff\ plot, with our calibration being given at a \logg\ of 4.44 dex (i.e. the Sun's surface gravity). The weighted reduced $\chi^{2}$ statistic is 0.14, 0.24, 1.55 and 0.46 for this work, Gray, Valenti \&\ Fischer and Bruntt respectively.

\begin{figure*}
\centering
\begin{minipage}{170mm}
\includegraphics[height=\columnwidth,angle=270]{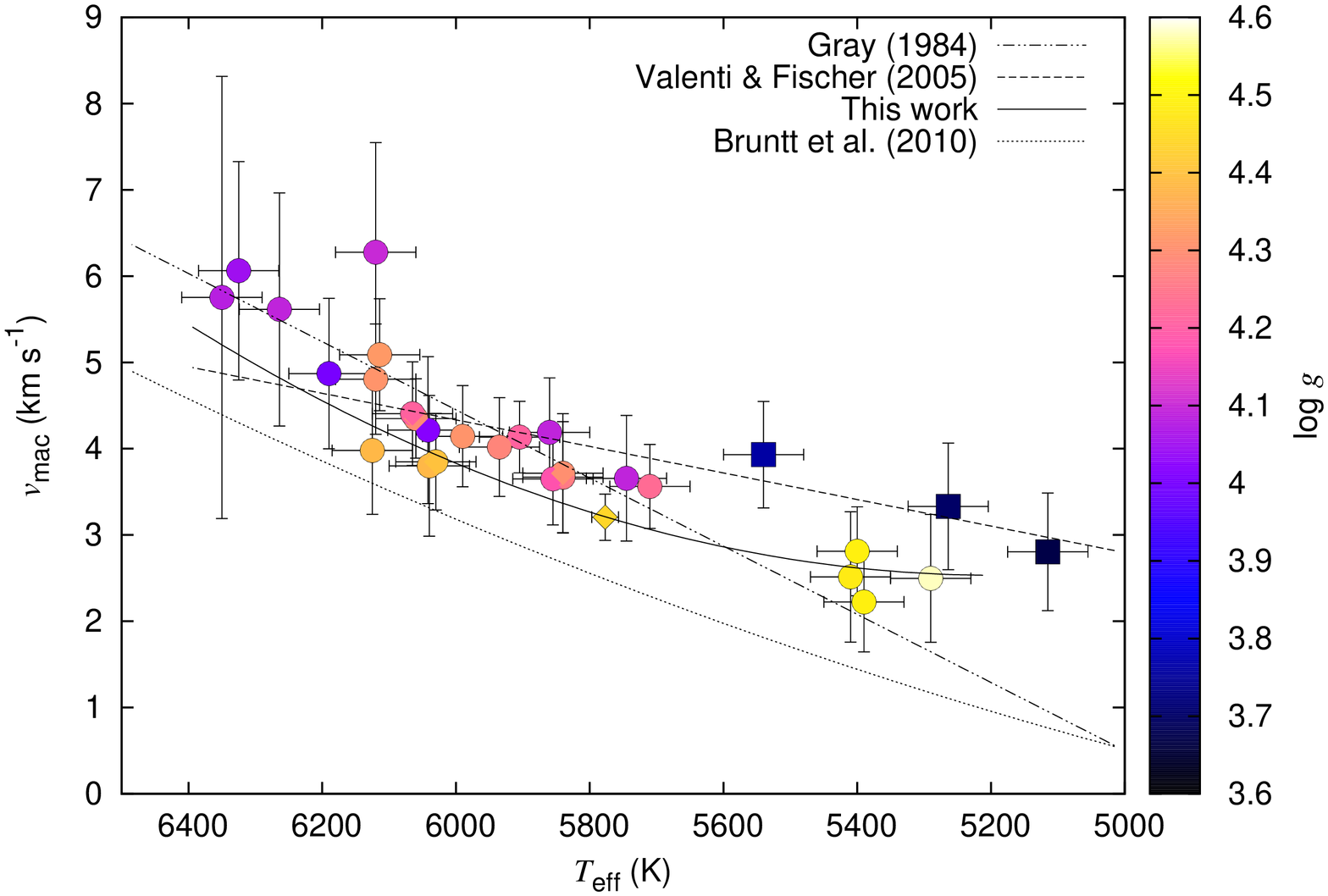}
\caption[]{Same as for Fig.~\ref{calibration}, but with the calibrations of \citet{Gray84},  \citet{Bruntt10a} and the upper limit of \citet{Valenti-Fischer05} also plotted. The fit for this work is given at the solar \logg\ = 4.44 dex.}
\label{compare}
\end{minipage}
\end{figure*}

\subsubsection{\citet{Gray84}}
\citet{Gray84} uses the Fourier method to determine the radial-tangential \mactrb\ for a selection of stars with high S/N (100:1 to 300:1) spectra. Gray stresses that there will be some uncertainties for late G and early K dwarfs, as the Zeeman broadening must be removed. Zeeman broadening was approximated as a convolution with the non-thermal profile, so that it could ultimately be removed.  \citet{Gray84} provides \mactrb\ values both with and without Zeeman broadening included in the profile, and there is a noticeable difference in the calibration in the cool end depending on which value is used. This might explain why our calibration results in higher \mactrb\ values for the cooler stars than Gray's calibration. The two calibrations show a similar trend for stars above 5700 K, showing the asteroseismic method is in agreement with the Fourier method. The differences are most likely due to the inclusion of \logg\ in our calibration and also because we used a polynomial instead of a linear fit. It should also be noted that Gray originally compared \mactrb\ to spectral type, and then converted spectral type to \teff.

\subsubsection{\citet{Valenti-Fischer05}}
\citet{Valenti-Fischer05} determined the radial-tangential \mactrb\ for each star by fitting a synthetic spectrum to an observed spectrum with the programme SME. They set \vsini\ = 0 \kms\ so the \mactrb\ derived is thus an upper limit because \mactrb\ was used to reproduce the effects of both rotation and macroturbulent broadening. Their Fig. 3 of the upper limit \mactrb\ plotted against \teff\ shows that below 5800 K, the slope changes by 1 \kms\ every 650 K. Using this, they then fit their linear relationship by fixing the solar \mactrb\ to 3.98 \kms\ (the value obtained by \citet{Gray84}) and the solar \vsini\ to 1.63 \kms\ \citep{Valenti-Piskunov96}. Their use of the solar \vsini\ of 1.63 \kms\ will result in a higher solar \mactrb\ than what we determined as we use a solar \vsini\ of 1.9 \kms.

They note that as stars below 5800 K should have negligible \vsini, then the resulting \mactrb\ is the true value rather than an upper limit. To determine if setting \vsini\ = 0 \kms\ has an effect on the \mactrb\ of stars below 5800 K, we compared \mactrb\ values obtained using \vsini\ = 0 to the \mactrb\ obtained when using the measured \vsini\ (0.7 $\textless$ \vsini\ $\textless$ 1.9 \kms) for the Sun and for the {\it Kepler} stars with \teff\ below 5800 K. The results show that using \vsini\ = 0 \kms\ for stars below 5800 K will overestimate the \mactrb\ by 0.41 $\pm$ 0.13 \kms, as can be seen in Fig.~\ref{vsini0}. Therefore, our results suggest that there may be a small systematic error in the \citet{Valenti-Fischer05} calibration for these stars. However from Fig.~\ref{compare} it is clear that it is still valid as an upper limit for dwarf stars below 5800 K. 

The solar \mactrb\ value of 3.98 \kms\ from \citet{Gray84} is that obtained solely from weak lines. These lines are formed deeper in the photosphere than strong lines, so that the weak lines will be prone to larger velocity fields. The value of 3.21 \kms\ determined from this work (see Section~\ref{mact-determination}) is an average value of both strong and weak lines and when this solar value is used in the \citet{Valenti-Fischer05} calibration, it is now in good agreement with Equation~\ref{fit} for stars with \teff\ $\textless$ 5800 K.

\begin{figure}
\centering
\includegraphics[height=\columnwidth,angle=270]{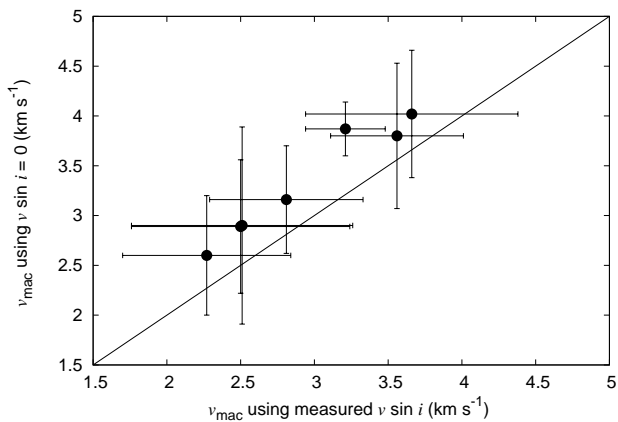}
\caption[]{Comparing the \mactrb\ obtained using the measured \vsini\ with the \mactrb\ obtained with \vsini\ = 0 \kms\ shows that the latter will be overestimated. The solid line depicts a 1:1 relationship. }
\label{vsini0}
\end{figure}

For stars with \teff\ greater than 5800 K, they note that the \mactrb\ points rise sharply above the linear relationship. They attribute this to the fact that \vsini\ will increase with \teff, but do acknowledge that some of this might be due to \mactrb. However their upper limit for \mactrb\ is no longer valid at these temperatures as the linear fit was determined for stars with \teff\ less than 5800 K.

\subsubsection{\citet{Bruntt10a}}
\citet{Bruntt10a} analysed a sample of stars using 10 to 30 lines with a line strength of between 20 and 100 m\AA\ using the software {\sc vwa} \citep{Bruntt02}. They used a \mactrb\ with a Gaussian profile, and convolved the synthetic spectrum with different combinations of \vsini\ and \mactrb\ in a grid with steps of 0.15 \kms\ until the best fit was found. A polynomial was fit to their data to determine their \mactrb\ calibration.

The Bruntt calibration gives \mactrb\ values that are systematically lower than ours. A line profile in {\sc vwa} with a given value of \mactrb\ is broader than in {\sc uclsyn}. In order for the two line profiles to agree, the \mactrb\ in {\sc vwa} needs to be increased by $\sqrt{2}$. For example, the solar \mactrb\ from the Bruntt calibration is 2.48 \kms, but when this is multiplied by $\sqrt{2}$ it gives 3.50 \kms, which is in agreement with \citet{Gray84}. This will also move the Bruntt calibration up on Figure~\ref{compare}. The $\sqrt{2}$ difference is probably due to the method of modelling \mactrb\ within the software, however the exact reason for this is unclear even to the authors of the code (Bruntt 2014, private communication).

If the \mactrb\ values are used as computed in \citet{Bruntt10a}, it would be expected that the \vsini\ will be pushed higher to compensate for this. In fact, this can be seen in Figure~\ref{Bruntt-vsini}, where the \vsini\ values from \citet{Bruntt12} are seen to be systematically higher than the $\vsini_{\rm astero}$ values. The stars that have a $\vsini_{\rm astero}$ too high to fit the spectra are also included on the plot.

\begin{figure}
\centering
\includegraphics[height=\columnwidth,angle=270]{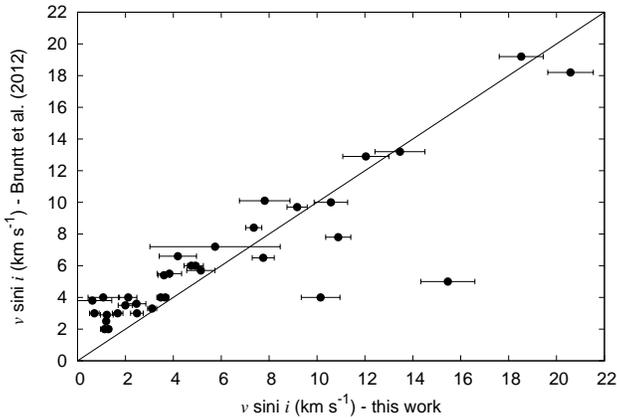}
\caption{Comparing the \citet{Bruntt12} \vsini\ values with the $\vsini_{\rm astero}$ values show that the Bruntt values are systematically higher. The outliers that have $\vsini_{\rm astero}$ too high for the line profiles are also shown on the plot. The solid line is a 1:1 relationship.}
\label{Bruntt-vsini}
\end{figure}

\subsection{Implications for the RM effect}

 \begin{table*}
\centering
\begin{minipage}{170mm}
\caption{The \mactrb\ has been calculated from equation~\ref{fit} and has an error of 0.73 \kms. The \vsini\ values redetermined for this work are given as $\vsini_{\rm spec}$, and the previous spectroscopic values ($\vsini_{\rm original}$) and the values obtained from the RM effect are also given. The \teff, \logg\ and \mictrb\ are determined from spectroscopic analyses from the given references.}
\begin{tabular}{l l l l l l l l l} \hline
Star & \teff\ & \logg\ & \mictrb\ & \mactrb & $\vsini_{\rm original}$ & $\vsini_{\rm RM}$ & $\vsini_{\rm spec}$ & References \\
     & K    &        & \kms\    & \kms\   & \kms                    & \kms\             & \kms\              &  \\ \hline
WASP-2  & 5175 $\pm$ 95 & 4.46 $\pm$ 0.12 & 0.70 $\pm$ 0.15 & 2.49 & 1.60 $\pm$ 0.70 & 0.99 $\substack{+0.27 \\ -0.32}$ & 0.88 $\pm$ 0.61 & 1, 2 \\
	& & & & &							1.30 $\pm$ 0.50 & \textless 0.5 & & 3 \\
WASP-4  & 5400 $\pm$ 90 & 4.47 $\pm$ 0.11 & 0.85 $\pm$ 0.10 & 2.56 & 2.00 $\pm$ 1.00 & 2.14 $\substack{+0.38 \\ -0.35}$ & 2.43 $\pm$ 0.37 & 1, 2 \\
WASP-5  & 5690 $\pm$ 80 & 4.28 $\pm$ 0.09 & 0.75 $\pm$ 0.10 & 3.34 & 3.50 $\pm$ 1.00 & 3.24 $\substack{+0.35 \\ -0.27}$ & 3.45 $\pm$ 0.37 & 1, 2 \\
WASP-6  & 5375 $\pm$ 65 & 4.61 $\pm$ 0.07 & 0.70 $\pm$ 0.10 & 2.26 & 1.40  $\pm$ 1.00 & 1.6 $\substack{+0.27 \\ -0.17}$ & 2.36 $\pm$ 0.31 & 1, 4 \\
WASP-8  & 5560 $\pm$ 90 & 4.40 $\pm$ 0.11 & 0.95 $\pm$ 0.15 & 2.88 & 2.00  $\pm$ 0.60 & 1.59  $\substack{+0.08 \\ -0.09}$ & 1.84 $\pm$ 0.38 & 1, 5 \\
WASP-15 & 6405 $\pm$ 80 & 4.40 $\pm$ 0.11 & 1.15 $\pm$ 0.08 & 5.54 & 4.00 $\pm$ 2.00 & 4.27 $\substack{+0.26 \\ -0.36}$ & 4.52 $\pm$ 0.46 & 1, 2 \\
WASP-16 & 5630 $\pm$ 70 & 4.21 $\pm$ 0.11 & 0.85 $\pm$ 0.10 & 3.37 & 2.3  $\pm$ 0.4 & 1.20  $\pm$ 0.3 & 1.90 $\pm$ 0.53 & 1, 6 \\
		& & & & &			 & 3.20  $\pm$ 0.90 & & 7 \\
WASP-18 & 6400 $\pm$ 75 & 4.32 $\pm$ 0.09 & 1.15 $\pm$ 0.08 & 5.68 & 11.00 $\pm$ 1.50 & 11.20 $\pm$ 0.60 & 10.96 $\pm$ 0.43 & 1, 2, 7 \\
WASP-19 & 5460 $\pm$ 90 & 4.37 $\pm$ 0.14 & 1.00 $\pm$ 0.15 & 2.81 & 5.0  $\pm$ 0.3 & 4.63  $\pm$ 0.26 & 4.86 $\pm$ 0.17 & 1, 8 \\
		& & & & & 4.0  $\pm$ 2.0 & 4.40  $\pm$ 0.90 & & 8 \\
WASP-20 & 6000 $\pm$ 100  & 4.40 $\pm$ 0.15 & 1.2 $\pm$ 0.1 & 3.91 & 3.5 $\pm$ 0.5 & 4.71 $\pm$ 0.50  & 3.92 $\pm$ 0.28 & 9 \\
WASP-21 & 5800 $\pm$ 100 & 4.2 $\pm$ 0.1 & 1.2 $\pm$ 0.1 & 3.57 & 1.5$\pm$ 0.6  & & 1.88 $\pm$ 0.42 & 10 \\
WASP-22 & 6000 $\pm$ 100 & 4.5 $\pm$ 0.2 & 1.2 $\pm$ 0.1 & 4.28 & 4.5  $\pm$ 0.4 & 4.42  $\pm$ 0.34 & 3.97 $\pm$ 0.30 & 11, 12 \\
WASP-24 & 6075 $\pm$ 100 & 4.15 $\pm$ 0.10 & 0.85 & 4.58 & 7.0 $\pm$ 1.0 & 7.32  $\pm$ 0.88 & 5.95 $\pm$ 0.28 & 13, 14 \\
WASP-25 & 5750 $\pm$ 100 & 4.5 $\pm$ 0.15 & 1.1 $\pm$ 0.1 & 3.03 & 2.6  $\pm$ 0.4 & 2.90  $\pm$ 0.3 & 2.35 $\pm$ 0.41 & 15, 6 \\
WASP-26 & 5950 $\pm$ 100 & 4.3 $\pm$ 0.2 & 1.2 $\pm$ 0.1 & 3.95 & 3.90 $\pm$ 0.4 &  2.20  $\pm$ 0.70  & 3.31 $\pm$ 0.31 & 7, 16, 12 \\
WASP-28 & 6100 $\pm$ 150 & 4.5 $\pm$ 0.2 & 1.2 $\pm$ 0.1 &  4.05 & 3.1 $\pm$ 0.6 & 3.25 $\pm$ 0.34 & 3.54 $\pm$ 0.49 & 9 \\
WASP-30 & 6190 $\pm$ 50 & 4.18 $\pm$ 0.08 & 1.1 $\pm$ 0.1 & 5.03 & 12.1 $\pm$ 0.5  & 12.1 $\substack{+0.4 \\ -0.5}$  & 11.84 $\pm$ 0.26 & 17 \\
WASP-31 & 6300 $\pm$ 100 & 4.4 $\pm$ 0.1 & 1.4 $\pm$ 0.1 & 5.23 & 7.6 $\pm$ 0.4 & 7.50  $\pm$ 0.7 & 7.56 $\pm$ 0.38 & 18, 6 \\
		& & & & &			 & 6.80  $\pm$ 0.60 & & 8 \\
WASP-32 & 6100 $\pm$ 100 & 4.4 $\pm$ 0.2 & 1.2 $\pm$ 0.1 & 4.25 & 5.5 $\pm$ 0.4 & 3.9  $\substack{+0.4 \\ -0.5}$ & 5.18 $\pm$ 0.27 & 19, 20 \\
WASP-38 & 6150 $\pm$ 80 & 4.3 $\pm$ 0.1 & 1.4 $\pm$ 0.1 & 4.64 & 8.3 $\pm$ 0.4 & 7.7  $\substack{+0.1 \\ -0.2}$ & 7.97 $\pm$ 0.25 & 21, 20 \\
		& & & & & 8.60 $\pm$ 0.40 & 			   & & 15 \\
WASP-40 & 5200 $\pm$ 150 & 4.5 $\pm$ 0.2 & 0.9 $\pm$ 0.2 & 2.41 & 2.4 $\pm$ 0.5 & 0.6  $\substack{+0.7 \\ -0.4}$ & 1.71 $\pm$ 0.39 & 22, 20 \\
WASP-41 & 5450 $\pm$ 100 & 4.4 $\pm$ 0.2 & 1.0 $\pm$ 0.2 & 2.74 & 1.6 $\pm$ 1.1 & & 2.74 $\pm$ 0.24 & 23 \\
WASP-50 & 5400 $\pm$ 100 & 4.5 $\pm$ 0.1 & 0.8 $\pm$ 0.2 & 2.50 & 2.6 $\pm$ 0.5 & & 2.65 $\pm$ 0.29 &  24 \\
WASP-54 & 6100 $\pm$ 100 & 4.2 $\pm$ 0.1 & 1.4 $\pm$ 0.2 & 4.65 & 4.0 $\pm$ 0.8 &  & 3.49 $\pm$ 0.42 &  25 \\
WASP-55 & 5900 $\pm$ 100 & 4.3 $\pm$ 0.1 & 1.1 $\pm$ 0.1 & 3.81 & 3.1 $\pm$ 1.0 &  & 2.42 $\pm$ 0.48 &  27 \\
WASP-61 & 6250 $\pm$ 150 & 4.3 $\pm$ 0.1 & 1.0 $\pm$ 0.2 & 5.04 & 10.3 $\pm$ 0.5 &  & 10.29 $\pm$ 0.36 & 26 \\
WASP-62 & 6230 $\pm$ 80 & 4.45 $\pm$ 0.10 & 1.25 $\pm$ 0.10 & 4.66 & 8.7 $\pm$ 0.4 & & 8.38 $\pm$ 0.35 & 26 \\
WASP-71 & 6050 $\pm$ 100 & 4.3 $\pm$ 0.1 & 1.4 $\pm$ 0.1 & 4.28 &  9.4 $\pm$ 0.5 & 9.89  $\pm$ 0.48 & 9.06 $\pm$ 0.36 & 27 \\
WASP-76 & 6250 $\pm$ 100 & 4.4 $\pm$ 0.1 & 1.4 $\pm$ 0.1 & 4.84 & 3.3 $\pm$ 0.6 & & 2.33 $\pm$ 0.36 & 28 \\
WASP-77A & 5500 $\pm$ 80 & 4.33 $\pm$ 0.08 & 0.8 $\pm$ 0.1 & 2.94 & 4.0 $\pm$ 0.2 & & 3.17 $\pm$ 0.34 & 29 \\
WASP-78 & 6100 $\pm$ 150 & 4.10 $\pm$ 0.20 & 1.1 $\pm$ 0.2 & 4.85 & 4.1 $\pm$ 0.2 & & 6.63 $\pm$ 0.16 & 30 \\

\hline
\end{tabular}
\label{HARPSvsini}
{\it References}: 1. \citet{Doyle13}, 2. \citet{Triaud10} , 3. \citet{Albrecht11}, 4. \citet{Gillon09}, 5. \citet{Queloz10}, 6. \citet{Brown12}, 7. \citet{Albrecht12}, 8. \citet{Hellier11}, 9. \citet{Anderson14}, 10. \citet{Bouchy10}, 11. \citet{Maxted10}, 12. \citet{Anderson11d}, 13. \citet{Street10}, 14. \citet{Simpson11}, 15. \citet{Enoch11b}, 16. \citet{Smalley10}, 17. \citet{Triaud13b}, 18. \citet{Anderson11c}, 19. \citet{Maxted10b}, 20.\citet{Brown12b}, 21. \citet{Barros11}, 22. \citet{Anderson11}, 23. \citet{Maxted11b}, 24. \citet{Gillon11}, 25. \citet{Faedi12}, 26. \citet{Hellier12}, 27. \citet{Smith13}, 28. \citet{West14}, 29.  \citet{Maxted13}, 30. \citet{Smalley12}.
\end{minipage}
\end{table*}

Using Equation~\ref{fit} to yield the \mactrb, we redetermined the \vsini\ for the WASP planet host stars using HARPS spectra. The resolution was determined individually for these spectra from the telluric lines, as the spectra were of a S/N high enough to do so. The typical resolution is $\sim$112\,000, in agreement with \citet{Mayor03}, where the resolution is given as 115\,000. The lines given in Table~\ref{mact-lines} were also used to fit \vsini. These new \vsini\ values are given in Table~\ref{HARPSvsini}, along with the original spectroscopic \vsini\ and the \vsini\ determined from the RM effect. The \mactrb\ values obtained will be affected by errors in the \teff\ and \logg. For instance, if the \teff\ of the WASP stars is increased by 100 K, the \mactrb\ will be 0.31 \kms\ greater on average. Similarly, an increase of 0.1 dex in \logg\ will increase the average \mactrb\ by 0.17 \kms.

A direct comparison with $\vsini_{\rm original}$ (and the RM values that used a spectroscopic prior) is difficult, as the \mactrb\ assumptions are inconsistent. However, there are some interesting comparisons to be made with the $\vsini_{\rm RM}$ values that did not require a spectroscopic prior. For example, $\vsini_{\rm spec}$ for WASP-40 seems to be higher than $\vsini_{\rm RM}$. The Zeeman effect can cause additional line broadening in stars cooler than G6 \citep{Gray84b}, so we checked to see if any such broadening was present using pairs of lines with high and low Land\'{e} g-factors as determined by \citet{Robinson80}. The lines should have similar depth of formation and line strength, so that when the \vsini\ is fixed to 1.71 \kms, the \mactrb\ measured from both lines should be the same if there is no additional broadening. In this sense, \mactrb\ was fitted to test for additional broadening of the lines, rather than to obtain the actual macroturbulent broadening.

If Zeeman broadening was affecting the line profiles, we would expect that the macroturbulence determined from the magnetically sensitive line would be higher than the reference line. WASP-40 does show evidence of Zeeman broadening, as seen in Figure~\ref{zeeman}, which shows the Fe line at 6842 \AA\ which has a  Land\'{e} g-factor of 2.5 overplotted with the Fe line at 6810 \AA\ which has lower Land\'{e} g-factor of 0.86. It is clear that the magnetically sensitive line at 6842 \AA\ exhibits stronger broadening. The \mactrb\ for this line is 3.15 \kms, while the \mactrb\ for the line at 6810 \AA\ is 2.10 \kms. This implies that $\vsini_{\rm spec}$ is overestimated for WASP-40 because the rotational broadening is used to erroneously fit also the Zeeman broadening. Therefore, if a spectroscopic prior is required for a cool star, it should be noted that if there is Zeeman broadening present, then the \vsini\ could be incorrect. There does not appear to be Zeeman broadening present in the cool {\it Kepler} stars in this study, however given the quality of the spectra it can not be completely ruled out.  

\begin{figure}
\centering
\includegraphics[height=\columnwidth,angle=270]{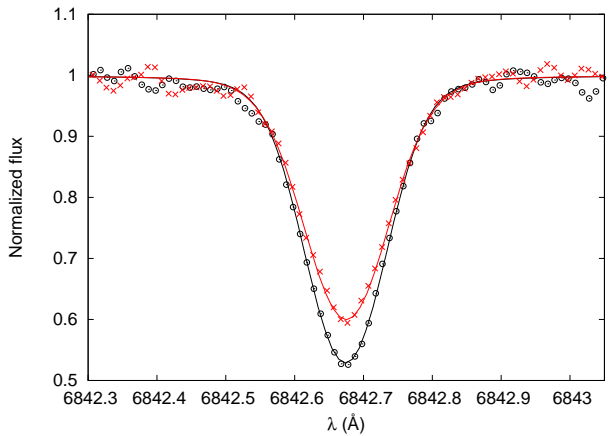}
\caption{The magnetically sensitive Fe line at 6842 \AA\ (red crosses) is overplotted with the non-magnetically sensitive Fe line at 6810 \AA\ (black circles) for WASP-40, showing that 6842 \AA\ exhibits stronger broadening which is likely due to Zeeman broadening. The solid lines are the synthetic fits to the observed spectrum.}
\label{zeeman}
\end{figure}

There are also some discrepancies for hotter stars, for example WASP-20, WASP-24 and WASP-32. In particular, $\vsini_{\rm RM}$ for WASP-24 is too high to fit the spectrum, which suggests that latitudinal differential rotation could be present. While it can be possible to detect differential rotation via the RM effect for misaligned planets that transit a range of latitudes \citep{Gaudi-Winn07}, these three systems are all well aligned. 

Overall, the use of Equation~\ref{fit} to determine \mactrb, and thus \vsini, has the advantage of having improved accuracy over the original spectroscopic values. However, they should still be used with caution as other factors can also influence the line broadening. 

\section{Conclusions and future work}
\label{conclusion}
We have used asteroseismic \vsini\ values obtained using {\it Kepler} to break the degeneracy between \vsini\ and \mactrb\ in spectral line profiles. By fixing the \vsini\ to the asteroseismic value, we were able to obtain the \mactrb\ for 28 {\it Kepler} stars. Out of this sample, 26 stars were used along with the Sun to derive a new calibration between between \mactrb, \teff, and \logg, which shows that there is an obvious trend between \mactrb\ and \teff, and also some indication of \logg\ dependence.

We used this calibration to determine \mactrb\ for some of the WASP planet host stars in a consistent manner, which enabled us to provide more accurate \vsini\ values.

The ESPaDOnS and Narval spectra used to measure the \mactrb\ in the {\it Kepler} stars are insufficiently sampled, and make it difficult to fit the line profiles. Higher S/N spectra are required in order to improve the calibration by enabling more precise continuum placement.

A number of the {\it Kepler} stars have asteroseismic \vsini\ values that are too high to fit the spectral lines. This suggests that latitudinal differential rotation might be present, which will be investigated in a forthcoming publication.

\section*{Acknowledgments}
We would like to thank Davide Gandolfi for his useful comments which helped to improve this paper. A.P.D. acknowledges support from EPSAM at Keele University. This research has made use of NASA's Astrophysics Data System and Ren\'{e} Heller's Holt-Rossiter-McLaughlin Encyclopaedia (www.physics.mcmaster.ca/$\sim$rheller). We would like to thank Sarbani Basu and David Brown for useful discussions, and Hans Bruntt for providing the ESPaDOnS and Narval spectra. Funding for the {\it Kepler} Discovery Class mission is provided by NASA's Science Mission Directorate. The authors wish to thank the entire {\it Kepler} team, without whom these results would not be possible. We would in particular like to thank R.A. Garc{\'i}a for preparation of the {\it Kepler} lightcurves for asteroseismic analysis. G.R.D.. and W.J.C. acknowledge the support of the UK Science and Technology Facilities Council (STFC). Funding for the Stellar Astrophysics Centre is provided by The Danish National Research Foundation (grant agreement No.: DNRF106).

%\bibliographystyle{mn2e_long}
%\bibliography{references}
%\include{bibliography}

\label{lastpage}

\end{document}